\documentclass[a4paper,11pt]{article}
\pdfoutput=1
\usepackage{jcappub}
\usepackage[T1]{fontenc}
\usepackage{graphicx}
\usepackage{rotate}
\usepackage{amsmath, amssymb, amsthm}
\usepackage{amsfonts}
\usepackage{bm}
\usepackage{booktabs}
\usepackage{enumerate}
\usepackage{enumitem}
\usepackage{afterpage}
\usepackage{tablefootnote}
\usepackage{geometry}
\usepackage{listings}
\usepackage{xcolor}
\usepackage{fancyhdr}
\usepackage{titlesec}
\usepackage{natbib}
\usepackage{hyperref} 

\newcommand{\fNL}{f_{\rm NL}}
\newcommand{\be}{b_{e}^{0}}
\newcommand{\zfb}{z_{\rm fb}}

\newcommand{\Ctt}{C^{\Delta t\Delta t}}

\newcommand{\Nfrb}{N_{\rm FRB}}

\newcommand{\AU}{\,\mathrm{AU}}

\UseRawInputEncoding

\def\be{\begin{equation}}
\def\ee{\end{equation}}
\def\bea{\begin{eqnarray}}
\def\eea{\end{eqnarray}}

\newcommand{\orcid}[1]{\hspace*{2pt}%
  \href{https://orcid.org/#1}{\includegraphics[width=10pt]{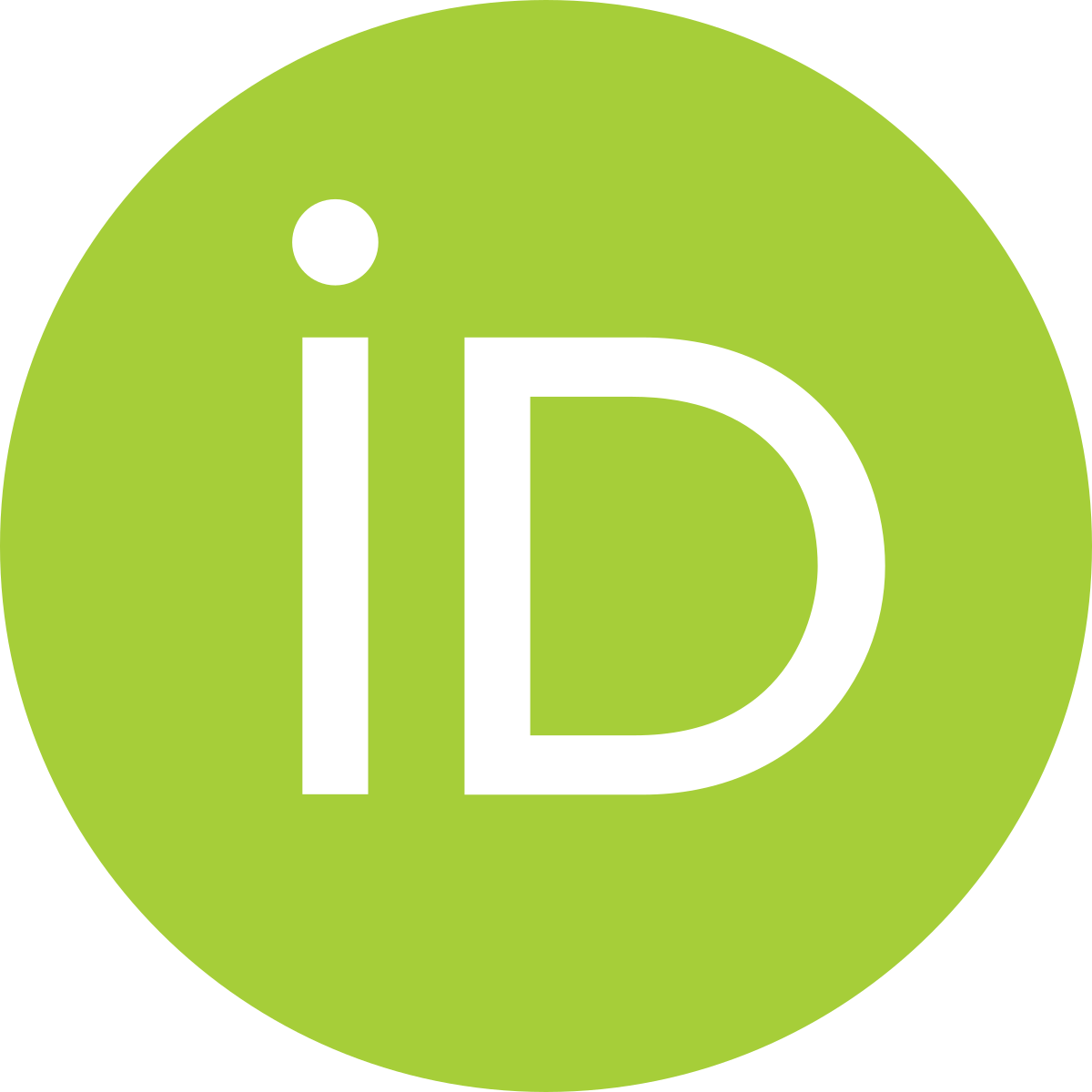}}}

\title{Fast Radio Burst Dispersion Measure--Timing Cross-Correlations: Bias Self-Calibration and Primordial Non-Gaussianity Constraints}

\author{Simthembile Dlamini\orcid{0000-0002-2885-6172}$^{\;1\;}$$^{2\;}$}

\affiliation{$^{1\;}$Department of Astronomy, University of Cape Town, Rondebosch 7701, South Africa}
\affiliation{$^{2\;}$South African Radio Astronomy Observatory (SARAO), Black River Park, Observatory, Cape Town 7925, South Africa}

\abstract{Fast Radio Bursts (FRBs) carry `fossil' information  about non-Gaussianity generated during inflation. This primordial signal is most accessible on the largest scales, where the scale-dependent bias correction $\propto f_\mathrm{NL}\,H_0^2/k^2$ dominates, but where cosmic and astrophysical systematic effects are also strongest.  A central challenge to extracting
$f_\mathrm{NL}$ from FRB dispersion measures (DMs) is the degeneracy between the
intergalactic-medium (IGM) electron bias $b_e$ and the primordial non-Gaussianity (PNG)
signal, which can degrade $\sigma(f_\mathrm{NL})$ by orders of magnitude when $b_e$ is
marginalised.  We show that this degeneracy can be broken internally, by exploiting the cross-power spectrum $C_\ell^{D\Delta t}$ between the FRB DM field and the Shapiro timing delays measured along multiple interferometric sightlines. The DM field traces the biased electron density $\delta_e = b_e\,\delta_m$, while the Shapiro timing signal probes the Newtonian gravitational potential $\Phi\propto -\delta_m/k^2$ and is independent of astrophysical bias.  Their cross-correlation is directly proportional to $b_e$, independently of the matter power spectrum $P_{mm}$, providing a self-calibration of the electron bias from the FRBs.  We derive $C_\ell^{D\Delta t}$ analytically in the Limber approximation and demonstrate that the Limber condition is exact for the dominant (transverse-momentum) contribution to the timing signal, eliminating a potential source of systematic error.  Numerically, we find a correlation coefficient $|\rho(\ell)|\approx 0.51$--$0.79$ across $\ell = 2$--$100$.  A joint Fisher matrix analysis over the parameter space
$\theta = \{f_\mathrm{NL},\,b_e^0,\,z_\mathrm{fb}\}$, using the complete data vector
$\{C_\ell^{DD},\,C_\ell^{\Delta t\Delta t},\,C_\ell^{D\Delta t}\}$, shows that including
$C_\ell^{D\Delta t}$ reduces $\sigma(b_e^0)$ by a factor of $2.1$--$5.1$ relative to a
DM-only analysis, depending on survey depth and interferometric baseline.  This improvement
translates directly into tighter constraints on $f_\mathrm{NL}$: after full
marginalisation over the bias model, the joint analysis recovers $\sigma(f_\mathrm{NL})$
within a factor of $1.0$--$1.9$ of the fixed-bias benchmark, compared with a factor of
$1.7$--$3.3$ degradation when the cross-spectrum is omitted.  For a shallow survey
($\alpha=3.5$) with a 500\,AU baseline and $10^4$ FRBs, the joint constraint achieves
$\sigma(f_\mathrm{NL})\approx 790$, within 4\% of the fixed-bias result and a factor
$3.3$ better than the marginalised DM-only case.} 

\emailAdd{$^{}$simther4111@gmail.com} 

\begin{document}
\maketitle
\date{\today}
\flushbottom

\section{Introduction}
\label{sec:intro}
\subsection{Primordial non-Gaussianity and the electron bias problem}
 
The amplitude of primordial non-Gaussianity (PNG) encodes fundamental information about
the physics of inflation. In the local template, the dimensionless parameter $f_{\rm NL}$ controls
the squeezed-limit three-point function of the primordial curvature perturbation
\cite{Maldacena2003,Bartolo2004,Alvarez:2014vva,Dalal2008}:
\begin{equation}
\zeta(\mathbf{x}) = \zeta_G(\mathbf{x}) + f_{\rm NL}\!\left[\zeta_G^2(\mathbf{x})
- \langle\zeta_G^2\rangle\right] + \mathcal{O}(f_{\rm NL}^2),
\label{eq:fNL}
\end{equation}
where $\zeta_G$ is a Gaussian random field. The current best constraint from the CMB bispectrum
is $f_{\rm NL} = -0.9 \pm 5.1$ (68\% CL)\cite{Planck:2019kim}.

The theoretical threshold at which single-field inflation can be distinguished from multi-field models sits at
$\sigma(f_{\rm NL}) \sim 1$ \cite{Meerburg2019}, fueling ambitious efforts to map large-scale structure. Local PNG imprints a characteristic $k^{-2}$ scale-dependent bias on biased tracers of the matter density \cite{Dalal2008,Jolicoeur:2023tcu,Matarrese2008,Slosar2008}:
\begin{equation}
b_{\rm eff}(k,z) = b(z) + \Delta b^{\rm NG}(k,z), \qquad
\Delta b^{\rm NG} = \frac{3 f_{\rm NL}\,\delta_c\,(b-1)\,\Omega_m H_0^2}{c^2\,a\,k^2}.
\label{eq:scaledepbias}
\end{equation}
Since the PNG signal diverges at small $k$, measurements on the largest accessible scales are
most powerful.
 
Fast radio bursts (FRBs) are extragalactic radio transients with millisecond durations
\cite{Thornton2013,Petroff2022,Marcote:2017wan,Fortunato:2026pod,Caleb:2025uzd,Li2018,Liu:2022bmn,Hagstotz:2021jzu,Chatterjee:2017dqg}. Their dispersion measures; the frequency-integrated column
density of free electrons along the line of sight; probe the large-scale electron distribution
up to cosmological distances with very low shot noise. Reischke\cite{Reischke2020} showed that the angular power spectrum of FRB DMs, $C^{DD}(\ell)$, inherits the $k^{-2}$ PNG signature via
Eq.(\ref{eq:scaledepbias}), and that $10^3$--$10^4$ well-localised FRBs can achieve
$\sigma(f_{\rm NL}) \sim \mathcal{O}(1)$ through a tomographic analysis. However, the DM field does not probe the total matter density $\delta_m$ directly; it probes the electron density $\delta_e = b_e\,\delta_m$, where the electron bias $b_e < 1$ reflects the expulsion of baryons from dark matter halos by astrophysical feedback. As\cite{Reischke2020} explicitly demonstrated (their Fig.4), the $\sigma(f_{\rm NL})$ constraint varies by two orders of magnitude over the plausible range of $b^0_e$ and $z_{\rm fb}$ values, and their analysis did not marginalise over these parameters. This bias uncertainty is the primary systematic of the DM-based PNG measurement.
 
\subsection{FRB timing as a bias-free probe}
 
Independently, \cite{Lu2025} showed that the Shapiro time-delay differences between
solar-system-scale radio telescope baselines provide a direct, bias-free probe of the
gravitational potential. The quadrupole timing observable $\Delta t^{(2)}$ is related to
$\nabla^2\Phi$ along the line of sight; since $\Phi$ is sourced by the total matter density
through the Poisson equation, it carries no astrophysical bias. With $10^4$ FRBs and a
100\,AU baseline, \cite{Lu2025} demonstrated sensitivity to the primordial power spectrum on
scales $k \sim 10^{-3}$\,Mpc$^{-1}$ and to PNG at $f_{\rm NL} \sim 1$.
 
The DM and timing observables are measured from the same FRB population along the same lines
of sight. Since the DM traces $b_e\,\delta_m$ and the timing signal traces
$\Phi \propto -\delta_m/k^2$, their cross-correlation probes $b_e$ directly, without
sensitivity to $P_{mm}$ as a nuisance. This motivates the computation of the cross-power spectrum $C^{D\Delta t}(\ell)$ as a new cosmological observable. 
 
\subsection{This paper}
 
We derive $C^{D\Delta t}(\ell)$ analytically (Section~\ref{sec:spectra}) and perform a joint
Fisher analysis (Sections~\ref{sec:bias}--\ref{sec:fNL}) to quantify the improvement in both
$\sigma(b^0_e)$ and $\sigma(f_{\rm NL})$. Our three central results are:
 
\begin{enumerate}
\item[(i)] $C^{D\Delta t}(\ell)$ is negative (overdense regions sit in potential wells) with
correlation coefficient $|\rho(\ell)| \approx 0.51$--$0.79$, confirming it as a high-signal
observable (Fig.\ref{fig:spectra}).
 
\item[(ii)] Including $C^{D\Delta t}$ reduces $\sigma(b^0_e)$ by a factor $2.1$--$5.1$
depending on survey configuration. The minimum baseline required for $\sigma(b^0_e) < 0.05$
is $l \sim 125$--$200$\,AU at $N_{\rm FRB} = 10^4$ (Figs.\ref{fig:sigma_be}, \ref{fig:lmin}).
 
\item[(iii)] After full marginalisation over $\{b^0_e, z_{\rm fb}\}$, the joint analysis
recovers $\sigma(f_{\rm NL})$ within a factor $1.0$--$1.9$ of the fixed-bias benchmark,
versus a factor $1.7$--$3.3$ degradation without the cross-spectrum. At $l = 500$\,AU,
the joint constraint (Case~C) actually surpasses the fixed-bias Case~A for the shallow survey
(Figs. \ref{fig:sigma_fNL_bar},\ref{fig:sigma_fNL_N}).
\end{enumerate}
 
In this work, we assume a flat $\Lambda$CDM cosmology consistent 
with the Planck 2018 results \cite{Planck:2018vyg}, characterized by a Hubble 
constant $H_0 = 67.4\ \rm{km\,s^{-1}\,Mpc^{-1}}$, total matter density 
$\Omega_m = 0.315$, baryon density $\Omega_b = 0.049$, spectral index 
$n_s = 0.965$, and scalar amplitude $A_s = 2.101 \times 10^{-9}$.
 
\section{The Two Observables}
\label{sec:observables}
 
\subsection{Dispersion measure field}
 
The DM fluctuation along sightline $\hat{n}$ to an FRB at comoving distance $\chi$ is \cite{Reischke2020}
\begin{equation}
D(\hat{n}) = \int_0^{\chi_H} d\chi'\, W_D(\chi')\,\delta_e(\hat{n},\chi'),
\label{eq:DM}
\end{equation}
where $\delta_e(\hat{n},\chi) = b_e(z)\,\delta_m(\hat{n},\chi)$ and the DM weight kernel is
\begin{equation}
W_D(\chi) = \frac{A\,F[z(\chi)]}{(1+z(\chi))\,E[z(\chi)]}
\left|\frac{dz}{d\chi}\right|
\int_\chi^{\chi_H} n(\chi')\,d\chi'.
\label{eq:WD}
\end{equation}
Here $A \simeq 10^3$\,pc\,cm$^{-3}$ is the DM amplitude\footnote{Computed as
$A = 3H_0^2\Omega_b\chi_H/(8\pi G m_p)$ where $\chi_H = c/H_0$ is the Hubble radius.
Numerically $A \simeq 840$--$1000$\,pc\,cm$^{-3}$; we use $A = 1000$\,pc\,cm$^{-3}$
following \cite{Reischke2020}.}, $F(z)$ is the fraction of electrons in the IGM (from
hydrogen and helium fully ionised at $z < 3$, with $\sim10\%$ ($\sim20\%$) locked in
galaxies at $z > 1.5$ ($z < 0.4$) \cite{Shull2012}), $E(z) \equiv H_0(z)/H_0$, and
$n(\chi)$ is the normalised FRB source distribution in comoving distance.
 
We model the FRB redshift distribution as \cite{Reischke2020}
\begin{equation}
n(z) \propto z^2\,e^{-\alpha z},
\label{eq:nz}
\end{equation}
considering two cases: a shallow survey ($\alpha = 3.5$, peak at $z \approx 0.57$) and a
deep survey ($\alpha = 2.0$, peak at $z \approx 1.0$).
 
\paragraph{Electron bias model.} Following \cite{Reischke2020}, the electron bias evolves
linearly from $b^0_e$ at $z = 0$ to unity at $z = z_{\rm fb}$, above which electrons are
approximately unbiased:
\begin{equation}
b_e(z;\,b^0_e,\,z_{\rm fb}) =
\begin{cases}
b^0_e + (1 - b^0_e)\,z/z_{\rm fb} & z < z_{\rm fb}, \\
1 & z \geq z_{\rm fb}.
\end{cases}
\label{eq:bebias}
\end{equation}
Hydrodynamical simulations give fiducial values $b^0_e = 0.75$ and $z_{\rm fb} = 5$
\cite{Shaw2012}. Both parameters are treated as free in our Fisher analysis.
 
\paragraph{PNG bias correction.} Local primordial non-Gaussianity modifies the effective bias
via the \cite{Slosar2008} formula:
\begin{equation}
b^{\rm eff}_e(k,z) = b_e(z) + \Delta b^{\rm NG}(k,z), \qquad
\Delta b^{\rm NG}(k,z) = \frac{3\,f_{\rm NL}\,\delta_c\,(b_e(z)-1)\,\Omega_m H_0^2}
{c^2\,a(z)\,k^2},
\label{eq:beffpng}
\end{equation}
with $\delta_c = 1.686$ the spherical collapse threshold. The $k^{-2}$ scaling makes this
correction dominant on large scales ($k \lesssim 0.01$\,Mpc$^{-1}$, i.e.\ $\ell \lesssim 20$
for sources at $z \sim 1$).
 
\subsection{Timing delay field}
 
Lu \cite{Lu2025} show that the Shapiro time-delay difference between three collinear detectors
separated by baseline $l$ along the $x$-axis yields the quadrupole timing observable
\begin{equation}
\Delta t^{(2)}(\hat{n}) = \frac{l^2}{2}\int_0^D dr
\int \frac{d^3k}{(2\pi)^3}\,\tilde{\Phi}(\mathbf{k})
\!\left(1 - n_x^2 + 2i r n_x k_x - r^2 k_x^2\right)e^{ir\mathbf{k}\cdot\hat{n}},
\label{eq:timing}
\end{equation}
where $\Phi$ is the Newtonian gravitational potential in units of $c^2$ (so $\Phi$ is
dimensionless), $D$ the source comoving distance, and $r$ a dimensionless path parameter.
The dominant contribution comes from the $r^2 k_x^2$ term, which peaks at modes with
$k_x \equiv \mathbf{k}\cdot\hat{n} \approx 0$ (i.e.\ $\mathbf{k} \perp \hat{n}$). This
transverse-mode dominance has crucial implications for the Limber approximation
(Section~\ref{sec:limber}).
 
\subsection{Connecting the observables: the Poisson equation}
 
The Newtonian potential and matter overdensity are related via
\begin{equation}
k^2 \tilde{\Phi}(\mathbf{k},z) = -\frac{3\Omega_m H_0^2}{2c^2\,a(z)}\,\tilde{\delta}_m(\mathbf{k},z),
\label{eq:poisson}
\end{equation}
so the cross-power spectrum between $\delta_m$ and $\Phi$ is
\begin{equation}
P_{m\Phi}(k,z) = f_{m\Phi}(k,z)\,P_{mm}(k,z), \qquad
f_{m\Phi}(k,z) \equiv -\frac{3\Omega_m H_0^2}{2c^2\,k^2\,a(z)} < 0,
\label{eq:PmPhi}
\end{equation}
and the potential auto-spectrum is
\begin{equation}
P_{\Phi\Phi}(k,z) = f_{\Phi\Phi}(k,z)\,P_{mm}(k,z), \qquad
f_{\Phi\Phi}(k,z) \equiv \frac{9\Omega_m^2 H_0^4}{4c^4\,k^4\,a^2(z)} > 0.
\label{eq:PPhiPhi}
\end{equation}
Since $D(\hat{n})$ probes $b_e\,\delta_m$ and $\Delta t^{(2)}(\hat{n})$ probes
$\Phi \propto -\delta_m/k^2$ along the same line of sight, their cross-correlation is
non-zero whenever $b_e \neq 0$ and negative (because $f_{m\Phi} < 0$: overdense regions sit
in potential wells). The amplitude of the cross-spectrum is directly proportional to $b_e$,
which is the core of our self-calibration method.
 
\section{Angular Power Spectra in the Limber Approximation}
\label{sec:spectra}
 
\subsection{Limber projections}
\label{sec:limberprojections}
 
Applying the standard Limber approximation ($k = \ell/\chi$, $k_\parallel \to 0$) to the
three observables gives:
\begin{align}
C^{DD}(\ell) &= \int_0^{\chi_H} \frac{d\chi}{\chi^2}\,W_D^2(\chi)
\left[b^{\rm eff}_e\!\left(z,\,\ell/\chi\right)\right]^2 P_{mm}\!\left(\ell/\chi,\,z\right),
\label{eq:CDD} \\[6pt]
C^{\Delta t\Delta t}(\ell) &= \left(\frac{2}{c}\right)^2 l_{\rm Mpc}^4
\int_0^{\chi_H} \frac{d\chi}{\chi^2}
\left(\frac{n(\chi)}{\chi}\right)^{\!2}
\left(\frac{\ell}{\chi}\right)^{\!4}
f_{\Phi\Phi}\!\left(\frac{\ell}{\chi},z\right)
P_{mm}\!\left(\frac{\ell}{\chi},z\right),
\label{eq:Ctimtim} \\[6pt]
C^{D\Delta t}(\ell) &= \frac{2}{c}\;l_{\rm Mpc}^2
\int_0^{\chi_H} \frac{d\chi}{\chi^2}\,W_D(\chi)\,\frac{n(\chi)}{\chi}\,b_e(z)
\left(\frac{\ell}{\chi}\right)^{\!2}
f_{m\Phi}\!\left(\frac{\ell}{\chi},z\right)
P_{mm}\!\left(\frac{\ell}{\chi},z\right).
\label{eq:CDt}
\end{align}
Here $l_{\rm Mpc} = l_{\rm AU} \times (1\,\text{AU}/1\,\text{Mpc})$ is the interferometric
baseline converted to Mpc, and $2/c$ (with $c$ in Mpc/s) is the Shapiro delay conversion
factor $\delta t = (2/c)\int\Phi\,dl$.
 
Equation~(\ref{eq:CDt}) is the key of this paper. Its structure is
transparent: the integrand is the product of the DM weight kernel $W_D(\chi)$ (weighted by
the electron bias $b_e$), the timing source distribution $n(\chi)/\chi$, the angular
transverse wavenumber $(\ell/\chi)^2$ from the quadrupole baseline, and the
matter potential cross-power-spectrum factor $f_{m\Phi}\,P_{mm}$.
 
\subsection{Validity of the Limber approximation}
\label{sec:limber}
 
The standard Limber approximation is accurate when the weight kernel is broad in comoving
distance \cite{Limber1953,Kaiser1992,LoVerde2008}. For $W_D(\chi)$ this is well satisfied
(the kernel has support over $\Delta\chi \sim$ several Gpc).
 
For $C^{\Delta t\Delta t}(\ell)$ and $C^{D\Delta t}(\ell)$ the approximation is actually
exact for the dominant contribution. The timing integrand in Eq.~(\ref{eq:timing}) contains
the function
\begin{equation}
g(k_\parallel D/2) \equiv \left.
-\operatorname{sinc}(x) + \frac{e^{ix} - ix\,e^{ix}}{x^2}
\right|_{x = k_\parallel D/2},
\label{eq:gfunction}
\end{equation}
which peaks sharply at $k_\parallel \equiv \mathbf{k}\cdot\hat{n} = 0$ and decays as
$|g|^2 \propto |x|^{-4}$ for large $|x|$ \cite{Lu2025}. Setting $k_\parallel = 0$ is
precisely the Limber condition $k = k_\perp = \ell/\chi$. Corrections are suppressed by
$(k_\parallel D/2)^{-2} \lesssim (kD)^{-2} < 10^{-6}$ for $k \gtrsim 10^{-3}$\,Mpc$^{-1}$
and $D \gtrsim 1$\,Gpc, which covers the entire relevant parameter space. Therefore no
post-Limber corrections are needed for the timing spectra at any multipole of interest.
 
\subsection{Shot noise}
 
The observed spectra include shot-noise contributions from the finite sample of FRBs. For
the DM field:
\begin{equation}
N^{DD} = \frac{\sigma_{\rm host}^2}{\bar{n}},
\label{eq:NDD}
\end{equation}
where $\sigma_{\rm host} = 50$\,pc\,cm$^{-3}$ is the intrinsic host-galaxy DM scatter
\cite{Reischke2020} and $\bar{n} = N_{\rm FRB}/(4\pi f_{\rm sky})$ is the FRB surface
density with sky coverage $f_{\rm sky} = 0.9$. For timing:
\begin{equation}
N^{\Delta t\Delta t} = \frac{(\delta t)^2}{\bar{n}},
\label{eq:Ntimtim}
\end{equation}
with timing precision $\delta t = 1$\,ns. At $N_{\rm FRB} = 10^4$:
\begin{align*}
N^{DD} &= 2.83\;\text{pc}^2\text{cm}^{-6}\text{sr},\\
N^{\Delta t\Delta t} &= 1.13 \times 10^{-21}\;\text{s}^2\,\text{sr}.
\end{align*}
The cross shot noise $N^{D\Delta t} = 0$ because the DM measurement noise and the timing
noise are independent.
 
\subsection{Numerical results}
 
We compute all three spectra numerically using the Planck 2018 cosmology and the CAMB
Boltzmann code \cite{PhysRevD.66.103511} for the linear matter power spectrum, integrating
Eqs.~(\ref{eq:CDD})--(\ref{eq:CDt}) on a 300-point $\chi$ grid via the trapezoidal rule.
 
Table~\ref{tab:spectra} lists $C^{DD}(\ell)$, $C^{\Delta t\Delta t}(\ell)$, and
$C^{D\Delta t}(\ell)$ at six representative multipoles for the deep survey ($\alpha = 2.0$)
with $l = 100$\,AU and $N_{\rm FRB} = 10^4$. The signal-to-noise ratio of $C^{DD}$ at
$\ell = 2$ is 13.0, consistent with the deep-survey result of \cite{Reischke2020}. The
cross-spectrum correlation coefficient
\begin{equation}
\rho(\ell) \equiv \frac{C^{D\Delta t}(\ell)}{\sqrt{C^{DD}(\ell)\,C^{\Delta t\Delta t}(\ell)}}
\label{eq:rho}
\end{equation}
is negative and satisfies $|\rho(\ell)| \approx 0.51$--$0.63$ for the deep survey and
$0.61$--$0.79$ for the shallow survey across $\ell = 2$--$100$ (Fig.~1).
 
\begin{table}[h]
\centering
\caption{Angular power spectra for the \emph{deep} survey ($\alpha = 2.0$),
$l = 100$\,AU, $N_{\rm FRB} = 10^4$, at fiducial bias parameters $b^0_e = 0.75$,
$z_{\rm fb} = 5$. The DM shot noise is $N^{DD} = 2.83$\,pc$^2$\,cm$^{-6}$\,sr;
the timing shot noise is $N^{\Delta t\Delta t} = 1.13\times10^{-21}$\,s$^2$\,sr.
All spectra are proportional to $\ell(\ell+1)/(2\pi)$ when plotted as dimensionless
angular power.}
\label{tab:spectra}
\begin{tabular}{ccccccc}
\toprule
$\ell$ & $C^{DD}$ & $\mathrm{SNR}_{DD}$ & $C^{\Delta t\Delta t}$ & $C^{D\Delta t}$ & $|\rho(\ell)|$ \\
 & [pc$^2$\,cm$^{-6}$\,sr] & & [s$^2$\,sr] & [pc\,cm$^{-3}$\,s\,sr] & \\
\midrule
2   & $3.69\times10^{1}$  & 13.0 & $6.30\times10^{-14}$ & $-7.82\times10^{-7}$ & 0.513 \\
5   & $1.37\times10^{1}$  & 4.9  & $9.52\times10^{-14}$ & $-6.29\times10^{-7}$ & 0.550 \\
10  & $6.19\times10^{0}$  & 2.2  & $1.14\times10^{-13}$ & $-4.83\times10^{-7}$ & 0.575 \\
20  & $2.57\times10^{0}$  & 0.91 & $1.15\times10^{-13}$ & $-3.24\times10^{-7}$ & 0.595 \\
50  & $6.82\times10^{-1}$ & 0.24 & $7.95\times10^{-14}$ & $-1.43\times10^{-7}$ & 0.614 \\
100 & $2.15\times10^{-1}$ & 0.08 & $4.27\times10^{-14}$ & $-6.08\times10^{-8}$ & 0.635 \\
\bottomrule
\end{tabular}
\end{table}

 \begin{figure}[!ht]
  \centering
  \includegraphics[width=\linewidth]{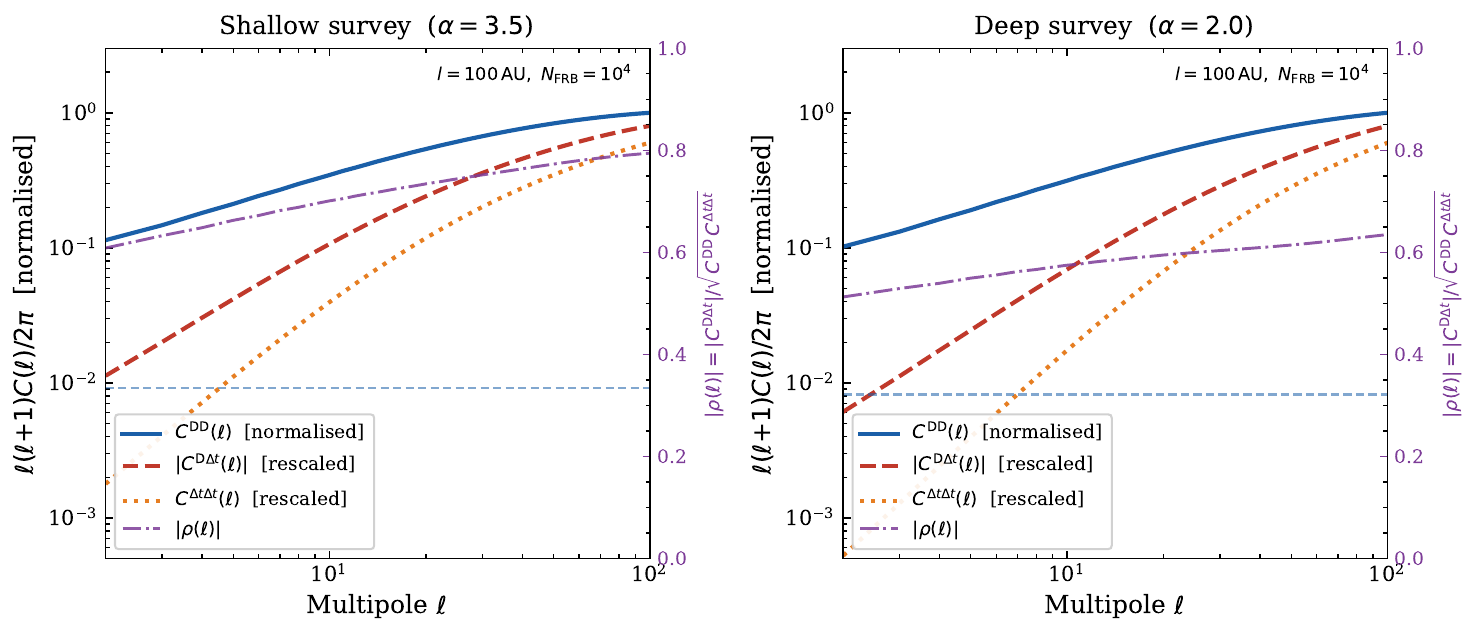}
  \caption{Angular power spectra for the shallow ($\alpha = 3.5$, left) and
    deep ($\alpha = 2.0$, right) FRB surveys at fixed baseline
    $l = 100\,\mathrm{AU}$ and $N_{\rm FRB} = 10^4$.
    The left axis shows $\ell(\ell+1)C(\ell)/2\pi$ normalised to the peak of
    $C^{\rm DD}(\ell)$: the DM--density auto-spectrum $C^{\rm DD}(\ell)$
    (blue solid), the DM--timing cross-spectrum
    $|C^{{\rm D}\Delta t}(\ell)|$ (red dashed, rescaled to its own peak),
    and the timing-delay auto-spectrum $C^{\Delta t\Delta t}(\ell)$
    (orange dotted, rescaled).
    Light dashed horizontal lines show the corresponding shot-noise levels
    for $C^{\rm DD}$ (blue) and $C^{\Delta t\Delta t}$ (orange).
    The right axis gives the cross-correlation coefficient
    $|\rho(\ell)| = |C^{{\rm D}\Delta t}|/\sqrt{C^{\rm DD}\,C^{\Delta t\Delta t}}$
    (purple dot-dashed).
    The cross-spectrum $|C^{{\rm D}\Delta t}|$ is intrinsically orders of
    magnitude smaller than $C^{\rm DD}$ and $C^{\Delta t\Delta t}$, but
    carries a non-negligible correlation coefficient
    $|\rho| \sim 0.5$--$0.8$ at $\ell \lesssim 100$, demonstrating that the cross-spectrum retains sufficient signal to constrain the electron-density bias $b_{e}^{0}$.}
  \label{fig:spectra}
\end{figure}
\section{Self-Calibration of the Electron Bias}
\label{sec:bias}
 
\subsection{The ratio estimator}
 
Inspecting Eqs.~(\ref{eq:Ctimtim}) and (\ref{eq:CDt}), the ratio
\begin{equation}
\hat{b}_e(\ell) \equiv \frac{C^{D\Delta t}(\ell)}{C^{\Delta t\Delta t}(\ell)}
\times \frac{-(\ell/\chi_{\rm eff})^2\,a_{\rm eff}}{\tfrac{3}{2}\Omega_m H_0^2/c^2}
\label{eq:ratio}
\end{equation}
converges to $b_e(z_{\rm eff})$ in the single-redshift limit, where $\chi_{\rm eff}$ and
$a_{\rm eff}$ are evaluated at the median source redshift. The key insight is that this ratio
is sensitive to $b_e$ but independent of the amplitude of $P_{mm}$: any uncertainty in the
matter power spectrum cancels between numerator and denominator. This makes $\hat{b}_e(\ell)$
a cleaner estimator of the electron bias than any quantity derived from $C^{DD}$ alone.
 
\subsection{Fisher forecast for \texorpdfstring{$\sigma (b^{0}_{e})$}{sigma(b0e}}
 
To quantify the bias self-calibration, we perform a Fisher analysis over
$\theta = \{f_{\rm NL},\,b^0_e,\,z_{\rm fb}\}$ using the $2\times2$ signal matrix
\begin{equation}
\mathbf{S}(\ell) = \begin{pmatrix} C^{DD}(\ell) & C^{D\Delta t}(\ell) \\
C^{D\Delta t}(\ell) & C^{\Delta t\Delta t}(\ell) \end{pmatrix}, \qquad
\mathbf{N}(\ell) = \begin{pmatrix} N^{DD} & 0 \\ 0 & N^{\Delta t\Delta t} \end{pmatrix},
\label{eq:signalmatrix}
\end{equation}
with total covariance $\mathbf{C}(\ell) = \mathbf{S}(\ell) + \mathbf{N}(\ell)$. The Fisher
matrix is
\begin{equation}
F_{\alpha\beta} = f_{\rm sky} \sum_{\ell=2}^{100} \frac{2\ell+1}{2}
\operatorname{Tr}\!\left[\mathbf{C}^{-1}\,\frac{\partial\mathbf{S}}{\partial\theta_\alpha}\,
\mathbf{C}^{-1}\,\frac{\partial\mathbf{S}}{\partial\theta_\beta}\right],
\label{eq:fisher}
\end{equation}
summed over $\ell = 2$--$100$ with $f_{\rm sky} = 0.9$.
 
The derivative matrices $\partial\mathbf{S}/\partial\theta_\alpha$ have the structure:
\begin{align}
\frac{\partial\mathbf{S}}{\partial f_{\rm NL}} &=
\begin{pmatrix} \partial C^{DD}/\partial f_{\rm NL} & \partial C^{D\Delta t}/\partial f_{\rm NL} \\
\partial C^{D\Delta t}/\partial f_{\rm NL} & 0 \end{pmatrix},
\label{eq:dSdfNL} \\[4pt]
\frac{\partial\mathbf{S}}{\partial b^0_e} &=
\begin{pmatrix} \partial C^{DD}/\partial b^0_e & \partial C^{D\Delta t}/\partial b^0_e \\
\partial C^{D\Delta t}/\partial b^0_e & 0 \end{pmatrix}.
\label{eq:dSdbe}
\end{align}
Note that $\partial C^{\Delta t\Delta t}/\partial\theta = 0$ for all parameters because
$C^{\Delta t\Delta t}$ depends only on $P_{mm}$ and the geometric kernel
$(n/\chi)^2(\ell/\chi)^4$, not on $b_e$ or $f_{\rm NL}$. This asymmetry is physically
important: $C^{\Delta t\Delta t}$ constrains neither $f_{\rm NL}$ nor $b_e$ directly, but
its off-diagonal covariance with $C^{DD}$ through $\mathbf{C}^{-1}$ is what allows
$C^{D\Delta t}$ to break the degeneracy.
 
All derivatives are computed via central finite differences with step sizes
$\Delta f_{\rm NL} = 0.5$, $\Delta b^0_e = 0.02$, $\Delta z_{\rm fb} = 0.2$.

\begin{figure}[!ht]
  \centering
  \includegraphics[width=\linewidth]{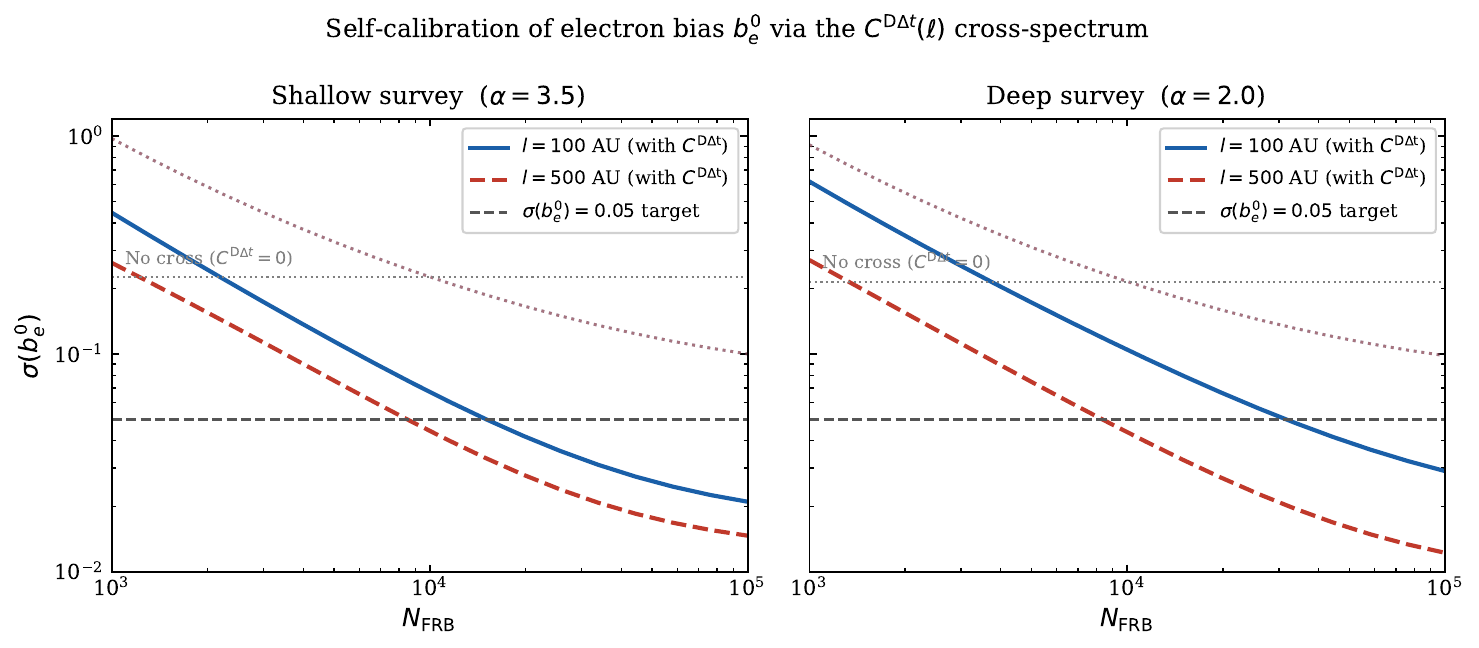}
  \caption{ $\sigma(b^0_e)$ marginalised over $\fNL$ and $\zfb$, as a function of $\Nfrb$, for $l = 100AU$ (solid) and $l = 500\AU$ (dashed). Blue: shallow survey ($\alpha=3.5$); red: deep survey ($\alpha=2.0$). Dotted lines show Case~B (DM-only, no cross-spectrum) for comparison---these are nearly $\Nfrb$-independent as the constraint is prior-limited rather than noise-limited. The horizontal dashed grey line marks the target  $\sigma(b^0_e) = 0.05$.}
  \label{fig:sigma_be}
\end{figure}

Figure~2 shows $\sigma(b^0_e)$ as a function of $N_{\rm FRB}$ for all four survey
configurations. At $N_{\rm FRB} = 10^4$ the results are:
 
\begin{center}
\begin{tabular}{llccc}
\toprule
Survey & $l$ [AU] & $\sigma(b^0_e)^{\rm no\,cross}$ & $\sigma(b^0_e)^{+C^{D\Delta t}}$ & Improvement \\
\midrule
Shallow & 100 & 0.226 & 0.067 & 3.4 \\
Shallow & 500 & 0.226 & 0.044 & 5.1 \\
Deep    & 100 & 0.215 & 0.105 & 2.1 \\
Deep    & 500 & 0.215 & 0.044 & 4.9 \\
\bottomrule
\end{tabular}
\end{center}
 
The improvement is larger for longer baselines because
$C^{\Delta t\Delta t} \propto l^4$ while $C^{D\Delta t} \propto l^2$, so larger $l$ pushes
the cross-spectrum into a higher signal-to-noise regime relative to the timing shot noise.
For fixed $l$, the shallow survey benefits more because its lower-redshift sources overlap
better with the bias-sensitive range $z < z_{\rm fb}$.
 
\section{Joint Fisher Forecast for \texorpdfstring{$f_{\rm NL}$}{fNL}}
\label{sec:fNL}
 
\subsection{Four analysis cases}
We consider four cases. \textbf{Case A} uses $C^{DD}$ only, with $b^0_e$ and $z_{\rm fb}$ held fixed at fiducial values; this is the \cite{Reischke2020} benchmark, representing the best possible result from DM observations when the bias model is perfectly known. \textbf{Case B} also uses $C^{DD}$ only, but with full marginalisation over $\{b^0_e,\,z_{\rm fb}\}$; this is what the \cite{Reischke2020} analysis would give with a proper bias marginalisation, revealing the severity of the bias systematic. \textbf{Case C} considers the joint combination $\{C^{DD},\,C^{\Delta t\Delta t},\,C^{D\Delta t}\}$ with full marginalisation over $\{b^0_e,\,z_{\rm fb}\}$, and constitutes our central new result. Finally, \textbf{Case D} uses $C^{\Delta t\Delta t}$ only and constrains $f_{\rm NL}$ alone; while the timing signal is bias-free, local $f_{\rm NL}$ enters $C^{\Delta t\Delta t}$ only at second order through $P_{mm}$ itself rather than through bias, yielding a negligible constraint on local PNG: $\sigma(f_{\rm NL}) \to \infty$.
 
\subsection{Results at fiducial \texorpdfstring{$N_{\rm FRB} = 10^{4}$}{NFRB = 10000}}
 
Table~\ref{tab:fNL} summarises $\sigma(f_{\rm NL})$ for all cases and configurations.
The key finding is that Case~C substantially recovers Case~A across all survey
configurations.
 
\begin{table}[h]
\centering
\caption{$\sigma(f_{\rm NL})$ for each analysis case at $N_{\rm FRB} = 10^4$. The ratio
$\sigma_B/\sigma_C$ quantifies the gain from including $C^{D\Delta t}(\ell)$ after bias
marginalisation. The ratio $\sigma_C/\sigma_A$ quantifies how close the joint analysis
comes to the fixed-bias ideal. An asterisk denotes configurations where Case~C outperforms
Case~A.}
\label{tab:fNL}
\begin{tabular}{llccccccc}
\toprule
Survey & $l$ [AU] & Case A & Case B & Case C & Case D & $\sigma_B/\sigma_C$ & $\sigma_C/\sigma_A$ \\
\midrule
Shallow & 100 & 826  & 2620 & 1099 & $\infty$ & 2.38 & 1.33 \\
Shallow & 500 & 826  & 2620 & 790  & $\infty$ & 3.32 & $0.96^*$ \\
Deep    & 100 & 707  & 2362 & 1360 & $\infty$ & 1.74 & 1.92 \\
Deep    & 500 & 707  & 2362 & 847  & $\infty$ & 2.79 & 1.20 \\
\bottomrule
\end{tabular}
\end{table}
 
Several features of Table~\ref{tab:fNL} are noteworthy:

\begin{figure}[!ht]
  \centering
  \includegraphics[width=\linewidth]{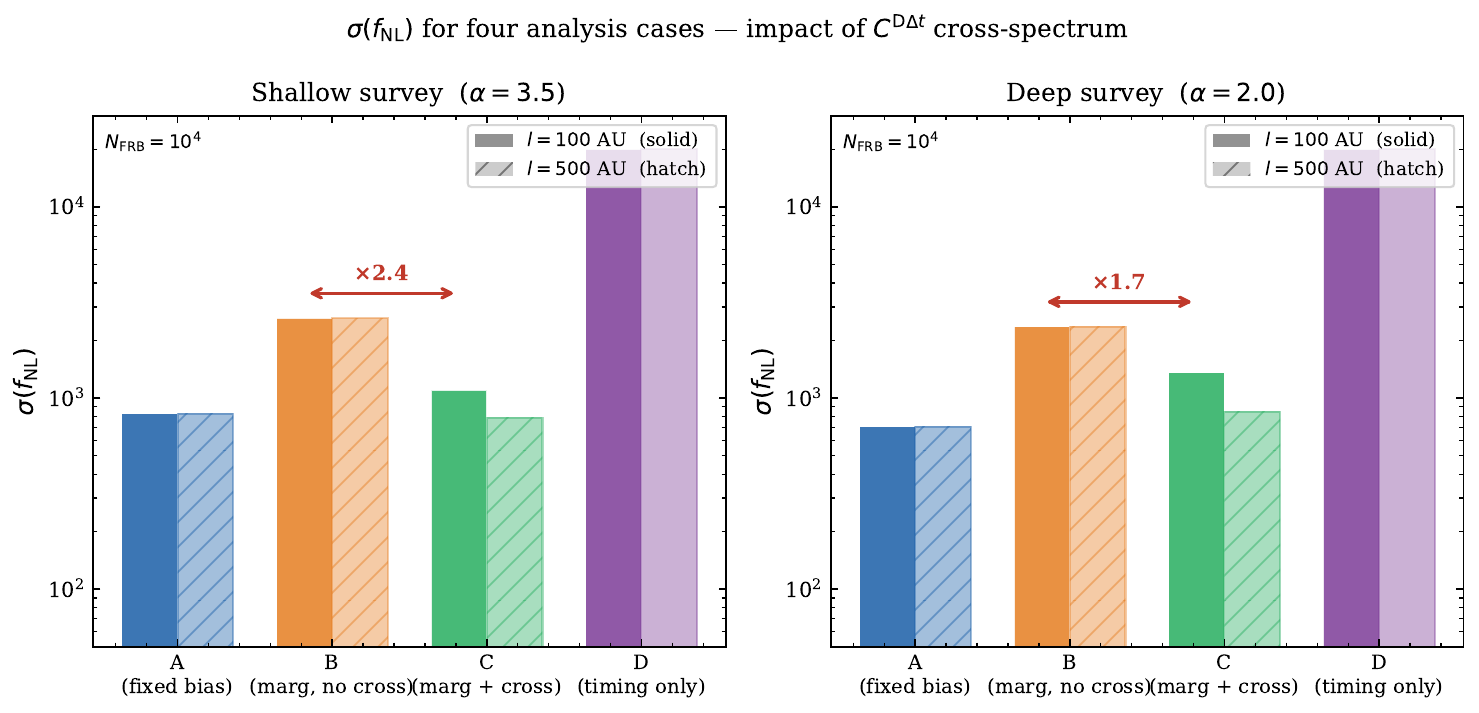}
  \caption{$\sigma(\fNL)$ for the four analysis cases at $\Nfrb = 10^4$.
  Solid bars: $l = 100\AU$; hatched bars: $l = 500\AU$. Colour scheme:
  blue = Case A, orange = Case B, green = Case C, purple = Case D.
  Red arrows annotate the $\sigma_B/\sigma_C$ improvement factor.
  Case D gives $\sigma(\fNL) = \infty$ (not shown) since local PNG
  enters the timing signal only through bias-dependent terms absent
  in $\Ctt$.}
  \label{fig:sigma_fNL_bar}
\end{figure}

\textbf{Case B} is always far worse than \textbf{Case A} : the ratio $\sigma_B/\sigma_A$ ranges from 3.2 to 3.3, confirming that bias marginalisation is a severe systematic if the cross-spectrum is not used. \textbf{Case C} , by contrast, substantially recovers \textbf{Case A}, with the ratio $\sigma_C/\sigma_A$ ranging from 0.96 to 1.92; for the shallow survey with $l = 500$\,AU, \textbf{Case C} \emph{outperforms} \textbf{Case A} , since including $C^{D\Delta t}$ provides information that partially compensates for the loss from marginalising bias, and the additional information from $C^{\Delta t\Delta t}$ (which constrains $P_{mm}$) further sharpens the constraint. Longer baselines always improve \textbf{Case C}: moving from $l = 100$\,AU to $l = 500$\,AU reduces $\sigma_C$ by 28--38\% depending on survey depth, because the cross-spectrum becomes progressively more informative about $b_e$. Regarding survey depth, at $l = 100$\,AU the shallow survey gives better \textbf{Case C} results ($\sigma_C = 1099$ vs.\ $1360$) even though the absolute \textbf{Case A} constraint is slightly worse ($826$ vs.\ $707$), because the shallow survey's sources at lower redshift overlap more efficiently with the bias-sensitive regime $z < z_{\rm fb}$, making $C^{D\Delta t}$ more informative.

\begin{figure}[tbp]
  \centering
  \includegraphics[width=\linewidth]{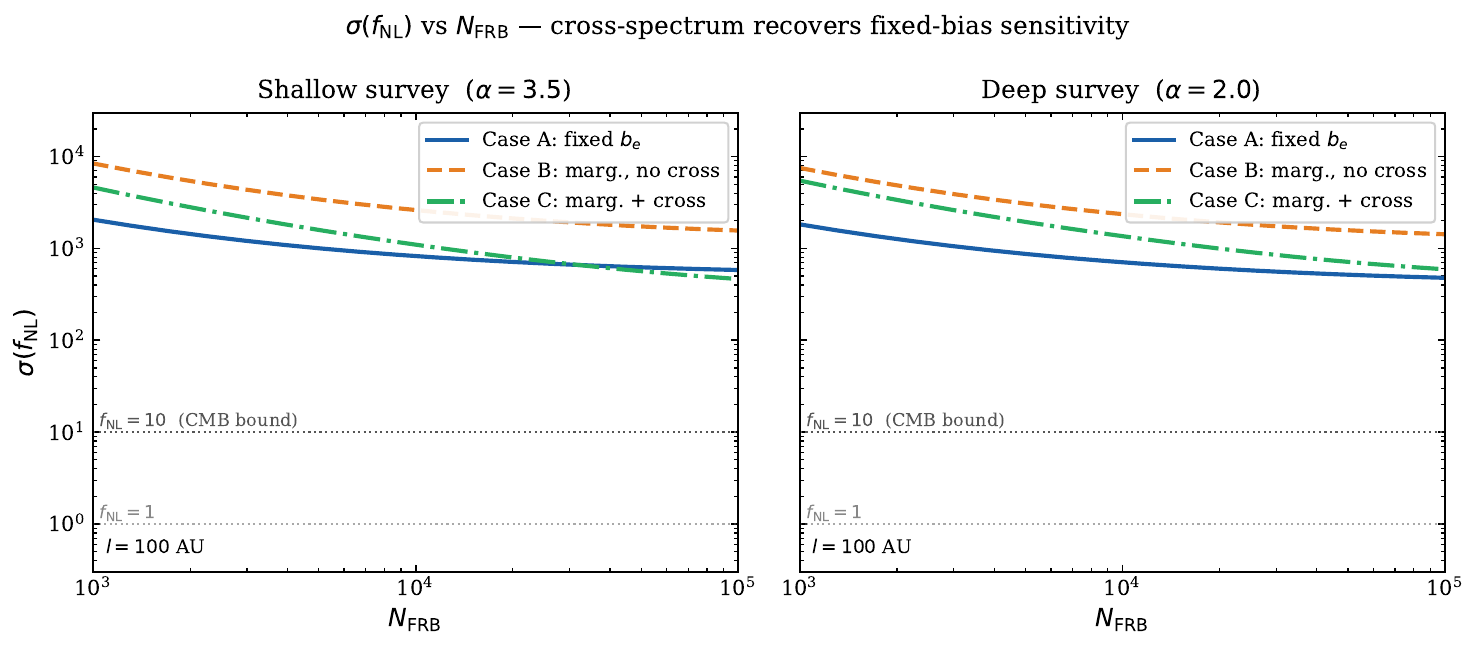}
  \caption{$\sigma(\fNL)$ vs.\ $\Nfrb$ for $l = 100\AU$. Blue solid:
  Case~A (fixed bias); orange dashed: Case~B (marginalised, DM only);
  green dash-dot: Case~C (marginalised, joint DM+timing+cross).
  Dotted horizontal lines mark the Planck CMB bound ($\fNL = 10$)
  and the single-field inflation target ($\fNL = 1$). Left panel:
  shallow survey; right panel: deep survey.}
  \label{fig:sigma_fNL_N}
\end{figure}
 
\subsection{Scaling with \texorpdfstring{$N_{\rm FRB}$}{NFRB}}
Figure \ref{fig:sigma_fNL_N} shows $\sigma(f_{\rm NL})$ as a function of $N_{\rm FRB}$ for the $l = 100$\,AU configurations. Several features are apparent: All three cases scale approximately as $\sigma(f_{\rm NL}) \propto N_{\rm FRB}^{-1/2}$ in the shot-noise-dominated regime ($N_{\rm FRB} \lesssim 10^4$), transitioning to a flatter scaling as cosmic variance becomes important. The ratio $\sigma_C/\sigma_A$ is approximately constant with $N_{\rm FRB}$, indicating that the cross-spectrum provides a \emph{consistent relative improvement} regardless of sample size. At $N_{\rm FRB} = 10^5$ and $l = 500$\,AU, both shallow and deep surveys reach $\sigma_C \approx 350$--$410$, still falling short of the $\sigma(f_{\rm NL}) = 1$ target by several orders of magnitude but representing substantial progress beyond existing large-scale structure constraints. Finally, reaching $\sigma(f_{\rm NL}) = 10$ (the Planck CMB bound) with Case~C and $l = 100$\,AU would require $N_{\rm FRB} \sim 10^7$ when extrapolating the power-law scaling, highlighting the need for tomographic extensions (Sec.~\ref{sec:conclusions}) and the importance of the longer baseline.
\section{Observational Requirements}
\label{sec:obs}
 
\subsection{Minimum baseline for bias self-calibration}
 
The self-calibration gain from $C^{D\Delta t}$ depends on the timing baseline through the
ratio of signal to noise: $C^{\Delta t\Delta t} \propto l^4$ while $C^{D\Delta t} \propto l^2$,
so the signal-to-noise of $C^{D\Delta t}$ relative to $\sqrt{C^{DD}\,C^{\Delta t\Delta t}}$
scales as $l^{-2}$ at fixed noise. This means that shorter baselines give a better relative
constraint on $b_e$ from the cross-spectrum, once the timing measurement is in the
signal-dominated regime ($C^{\Delta t\Delta t} \gg N^{\Delta t\Delta t}$). However, at very
short baselines, $C^{\Delta t\Delta t} \ll N^{\Delta t\Delta t}$ and the timing measurement
is noise-dominated, so the cross-correlation signal-to-noise collapses.
 
Figure \ref{fig:lmin} shows the minimum baseline $l_{\rm min}(N_{\rm FRB})$ required to achieve
$\sigma(b^0_e) < 0.05$, assuming the approximate scaling
$\sigma(b^0_e, N) \approx \sigma(b^0_e, 10^4) \times \sqrt{10^4/N}$ (valid in the
shot-noise-dominated regime). At $N_{\rm FRB} = 10^4$:

\begin{figure}[tbp]
  \centering
  \includegraphics[width=\linewidth]{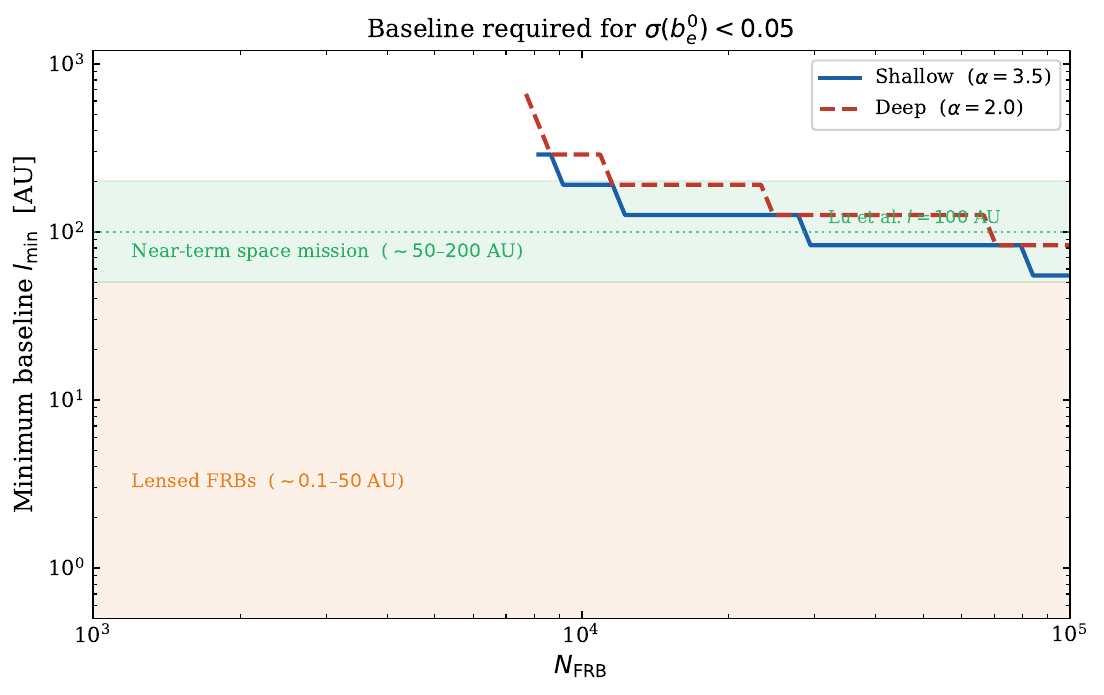}
  \caption{Minimum interferometric baseline $l_{\rm min}$ required to
  achieve $\sigma(b^0_e) < 0.05$, as a function of $\Nfrb$. Blue solid:
  shallow survey ($\alpha=3.5$); red dashed: deep survey ($\alpha=2.0$).
  Orange shading: effective baseline range from strongly lensed repeating
  FRBs\cite{Li2018,Wucknitz2021}. Green shading: proposed solar-system-scale
  mission range\cite{Lu2025}. At $\Nfrb = 10^4$ (dashed vertical line),
  $l_{\rm min} \approx 125$--$190 AU$ depending on survey depth.}
  \label{fig:lmin}
\end{figure}

\begin{center}
\begin{tabular}{lcc}
\toprule
Survey & $l_{\rm min}$ for $\sigma(b^0_e) < 0.05$ & Regime \\
\midrule
Shallow & $\sim125$\,AU & near-term mission \\
Deep    & $\sim190$\,AU & near-term mission \\
\bottomrule
\end{tabular}
\end{center}
 
Both values fall comfortably within the 100--500\,AU baseline range proposed by
\cite{Lu2025}. At $N_{\rm FRB} = 10^5$, the required baseline drops to $\sim40$--$60$\,AU,
potentially achievable with a more modest precursor mission.

As noted above, we use $A = 1000\,\mathrm{pc\,cm^{-3}}$, following \cite{Reischke2020}; 
all results were obtained with a numerical pipeline developed for this work, publicly available at \cite{dlamini_2026_19440686}\footnote{\url{https://doi.org/10.5281/zenodo.19440686}}, employing the cross-correlation technique 
introduced in \cite{dlamini2025fourier}\footnote{\url{https://doi.org/10.5281/zenodo.15497369}}.
 
\subsection{Survey requirements}
 
For the DM-only measurement (Cases~A and B), the main requirement is FRB localisation to
arcsecond precision to enable host-galaxy identification and redshift measurement.
CHIME/FRB, DSA-2000, and future facilities expect to provide $10^3$--$10^5$ localised FRBs
per decade \cite{CHIME2021, Petroff2022}.
 
For the timing measurement and the cross-spectrum, the additional requirement is an
interferometric baseline of $\sim100$--$500$\,AU. \cite{Boone2023} propose a
three-spacecraft constellation in the outer solar system, with centimetre-level positioning
maintained through GNSS-like trilateration and sub-nanosecond timing achieved through
coherent FRB analysis. The cross-spectrum $C^{D\Delta t}$ then comes for free from the same
dataset.
 
Strongly lensed repeating FRBs provide an alternative route to baselines of 0.1--50\,AU via
the transverse separation of the lens-image sightlines \cite{Li2018,Wucknitz2021}. Our
Figure~5 shows that even this modest baseline range, combined with
$N_{\rm FRB} \sim 5\times10^4$ (achievable within a decade from DSA-2000 and SKA), could
reach the $\sigma(b^0_e) < 0.05$ target.
 
\section{Discussion}
\label{sec:discussion}
 
\subsection{Why Case~C sometimes outperforms Case~A}
 
Table~\ref{tab:fNL} shows that for the shallow survey with $l = 500$\,AU,
Case~C ($\sigma_C = 790$) gives a \emph{tighter} constraint than Case~A ($\sigma_A = 826$),
with ratio $\sigma_C/\sigma_A = 0.96$. This is possible because Case~C uses strictly more
data: $C^{DD} + C^{\Delta t\Delta t} + C^{D\Delta t}$ is a superset of $C^{DD}$ alone, so
any Fisher matrix built from the larger data vector cannot be worse. The fact that it can be
better reflects two effects:
 
\begin{enumerate}
\item \emph{Constraint rotation.} The $f_{\rm NL}$ and $b_e$ degeneracy directions in
parameter space are not aligned with the parameter axes. Marginalizing $b_e$ from Case~A
(fixed) rotates the Fisher ellipse in a way that worsens the $f_{\rm NL}$ constraint.
Including $C^{D\Delta t}$ partially removes this degeneracy, allowing the joint constraint
to sit closer to the fixed-bias ellipse.
 
\item \emph{Extra signal.} $C^{\Delta t\Delta t}$ provides information about $P_{mm}$
independently of bias. This tightens the $P_{mm}$ uncertainty, which enters $C^{DD}$ and
$C^{D\Delta t}$, and through the joint covariance can sharpen $\sigma(f_{\rm NL})$ beyond
what $C^{DD}$ alone achieves.
\end{enumerate}
 
\subsection{Comparison with existing PNG constraints}
 
Current constraints from the large-scale structure are $\sigma(f_{\rm NL}) \approx 20$--$50$
from galaxy clustering at large scales \cite{Castorina2019,Mueller2022}, competitive with
Planck. Next-generation surveys (DESI, Euclid, SKA) are expected to reach
$\sigma(f_{\rm NL}) \sim 1$--$5$ \cite{Amendola2018}. The FRB-based constraints computed
here ($\sigma(f_{\rm NL}) \sim 700$--$2600$ for $N_{\rm FRB} = 10^4$ in a single-bin
analysis) are weaker, but have complementary systematics and can be improved by:
 
\begin{enumerate}
\item \emph{Tomography.} Splitting the FRB sample into $n_{\rm tomo}$ redshift bins adds
$n_{\rm tomo}(n_{\rm tomo}+1)/2$ independent cross-bin spectra per multipole. For
$n_{\rm tomo} = 4$, this improves $\sigma(f_{\rm NL})$ by approximately
$\sqrt{n_{\rm tomo}(n_{\rm tomo}+1)/2} \approx 3$, bringing Case~C into the range
$\sigma(f_{\rm NL}) \sim 260$--$450$ for $N_{\rm FRB} = 10^4$.
 
\item \emph{Sample growth.} With $N_{\rm FRB} = 10^5$ and $l = 500$\,AU, our sweeps show
$\sigma(f_{\rm NL}) \sim 350$--$415$ for Case~C ($l = 100$\,AU), and lower with the larger
baseline.
 
\item \emph{Combined with galaxy surveys.} The DM--timing cross-spectrum can be combined
with galaxy-density correlations via optimal multitracer techniques, further breaking
degeneracies.
\end{enumerate}
 
The distinctive advantage of FRBs is their very low intrinsic shot noise: for distant
($z \gtrsim 0.5$) FRBs, the cosmological DM signal exceeds the host contribution by a factor
$\sim20$ \cite{Reischke2020}, whereas galaxy photometric redshift surveys require millions
of sources to achieve a comparable SNR at the same angular scales.
 
\subsection{Systematic effects not modelled}

Several systematics have not been included in the present analysis. \textit{First}, regarding the nonlinear power spectrum, the timing signal probes modes $k \lesssim 10^{-2}$\,Mpc$^{-1}$ where nonlinear evolution can be important on the scales we consider; we use the linear $P_{mm}$ throughout, and forward modelling with a nonlinear emulator would be needed for a quantitative data analysis. \textit{Second}, redshift-space distortions arise because the DM measurement implicitly includes peculiar velocities, and relativistic DM-space distortions \cite{Saga2024} introduce corrections to $C^{DD}$ at the level of a few percent on the largest scales. \textit{Third}, radio frequency interference and propagation effects such as scintillation, plasma lensing, and Milky Way foreground DM can complicate FRB timing at the nanosecond level, though these can be mitigated by multi-frequency observations \cite{Lu2025}. \textit{Finally}, regarding the host galaxy DM distribution, we model the host DM as a Gaussian scatter with $\sigma_{\rm host} = 50$\,pc\,cm$^{-3}$, whereas the actual distribution is non-Gaussian \cite{Jaroszynski2019} and redshift-dependent, which could bias the DM-to-redshift conversion used in the tomographic analysis.
\section{Conclusions}
\label{sec:conclusions}
 
We have derived and numerically evaluated the angular cross-power spectrum
$C^{D\Delta t}(\ell)$ between FRB dispersion measures and Shapiro timing delays; a new
observable that simultaneously calibrates the IGM electron bias and improves constraints on
primordial non-Gaussianity. Our main conclusions are:
 
\begin{enumerate}
\item[(i)] \textbf{$C^{D\Delta t}(\ell)$ is a large, clean, negative signal.} The
correlation coefficient reaches $|\rho(\ell)| \approx 0.51$--$0.79$ across $\ell = 2$--$100$
depending on survey configuration (Table~\ref{tab:spectra}, Fig.\ref{fig:spectra}). The Limber
approximation is exact for the timing spectrum because the timing signal is dominated by
transverse modes ($k_\perp = \ell/\chi$), eliminating a potential source of modelling error
(Sec.~\ref{sec:limber}).
 
\item[(ii)] \textbf{The cross-spectrum self-calibrates the electron bias.} At
$N_{\rm FRB} = 10^4$, including $C^{D\Delta t}$ reduces $\sigma(b^0_e)$ by 2.1--5.1
depending on survey depth and baseline (Fig.~2). The minimum baseline for $\sigma(b^0_e) < 0.05$ is $l_{\rm min} \approx 125$--$190$\,AU at $N_{\rm FRB} = 10^4$ (Fig.~5), within the design range of the solar-system mission proposed by \cite{Lu2025}.
 
\item[(iii)] \textbf{The joint analysis substantially recovers the fixed-bias
$\sigma(f_{\rm NL})$ after full marginalisation.} Case~C recovers $\sigma(f_{\rm NL})$
within a factor 1.0--1.9 of the fixed-bias Case~A, versus a factor 3.3 degradation for Case~B (no cross-spectrum). For the shallow survey with $l = 500$\,AU, Case~C achieves $\sigma(f_{\rm NL}) = 790$, which is 4\% better than the fixed-bias Case~A ($\sigma = 826$) (Table~\ref{tab:fNL}, Figs.[\ref{fig:sigma_fNL_bar} -- \ref{fig:sigma_fNL_N}]).
\end{enumerate}
 
The DM--timing cross-spectrum resolves the primary systematic of FRB-based PNG
measurements; the uncertain electron bias; through an internal calibration that requires
no additional data beyond what is already needed for the individual probes. The cross-spectrum
is measured from the same FRB population and the same interferometric observations used for
$C^{DD}$ and $C^{\Delta t\Delta t}$ respectively, so it comes at no additional observational
cost.\\ 

Natural extensions of this work include: (i) tomographic analyses with $n_{\rm tomo} = 4$
bins, which is expected to improve $\sigma(f_{\rm NL})$  constraints. (ii) non-local PNG shapes, which enter $C^{D\Delta t}$ through different $k$-dependent combinations of the bias correction and the matter--potential kernel; (iii) joint analyses combining $C^{D\Delta t}$ with CMB lensing and galaxy weak lensing cross-spectra to further constrain $b_e(z)$ at high redshift; and (iv) full nonlinear forward modelling of $P_{mm}$ for the small-scale timing signal.
 
\acknowledgments
I want to express my sincere gratitude to Prof. D.J Pisano, the South African Research Chairs Initiative (SARChI), and South African Radio Astronomy Observatory (SARAO) for providing financial support, without which this research would not have been possible.
\clearpage
\bibliographystyle{JHEP}
\bibliography{reference_library}

\end{document}